\documentclass[aps,pra,floats,floatfix,showpacs,showkeys,twocolumn,superscriptaddress]{revtex4}
\usepackage{amssymb}
\usepackage{epsfig}
\usepackage{graphicx}
\usepackage{amsmath}

\begin{document}

\title{Ultrafast stimulated Raman parallel adiabatic passage by shaped pulses}

\begin{abstract}
We present a general and versatile technique of population
transfer based on {\it parallel adiabatic passage} by femtosecond
shaped pulses. Their amplitude and phase are specifically designed
to optimize the adiabatic passage corresponding to parallel
eigenvalues at all times. We show that this technique allows the
robust adiabatic population transfer in a Raman system with the
total pulse area as low as 3 $\pi$, corresponding to a fluence of
one order of magnitude below the conventional stimulated Raman
adiabatic passage process. This process of short duration,
typically pico- and subpicosecond, is easily implementable with
the modern pulse shaper technology and opens the possibility of
ultrafast robust population transfer with interesting applications
in quantum information processing.
\end{abstract}

\author{G. Dridi}
 \affiliation{Institut Carnot de
Bourgogne UMR 5209 CNRS, Universit\'e de Bourgogne, BP 47870,
21078 Dijon, France} \affiliation{Facult\'e des Sciences de Tunis
(FST)- D\'epartement de physique. Campus Universitaire El Manar,
1060 Tunis, Tunisia}
\author{S.~Gu\'erin}
\email{sguerin@u-bourgogne.fr} \affiliation{Institut Carnot de
Bourgogne UMR 5209 CNRS, Universit\'e de Bourgogne, BP 47870,
21078 Dijon, France}
\author{V. Hakobyan}
\affiliation{Institut Carnot de Bourgogne UMR 5209 CNRS,
Universit\'e de Bourgogne, BP 47870, 21078 Dijon, France}
\author{H.R~Jauslin}
\affiliation{Institut Carnot de Bourgogne UMR 5209 CNRS,
Universit\'e de Bourgogne, BP 47870, 21078 Dijon, France}
\author{H. Eleuch}
%\altaffiliation{The Abdus Salam Centre for Theoretical Physics,
%ICTP, Strada Costiera 11, 34014 Trieste, Italy}
\affiliation{Institut National des Sciences Appliqu\'ees et de
Technologie, BP 676, Zone Urbaine Nord, 1080 Tunis, Tunisia}
\date{\today }
\maketitle

\section{Introduction}
Complete or partial population transfer between ground states by
pulsed laser fields is at the heart of many quantum processes
\cite{Shore} ranging from laser-controlled chemistry
\cite{Shapiro} to modern quantum optics and quantum information,
such as quantum gates \cite{QI}.

Standard $\pi$-pulse techniques that require a specific pulse area
are known to lack robustness \cite{Allen,Holthaus}. Adiabatic
passage, which allows the dynamics to follow a single eigenstate
of the system, is often used to implement such transfers due to
its robustness with respect to field fluctuations and to the
imperfect knowledge of the studied system. In particular, the
stimulated Raman adiabatic passage (STIRAP) process, which induces
a complete population transfer in a $\Lambda$ system
\cite{STIRAP}, and various extensions
\cite{Vitanov-review,Shapiro_CC,Adv},
%in particular to generate
%superpositions of states \cite{fstirap} and gates
%\cite{Kis,Lacour,Shapiro_Gate}
have become very popular. The main advantage relies on its
dark-state dynamics which does not involve the upper state in the
adiabatic limit, and makes thus this process in principle immune
to decoherence. However this process requires in practice a large
pulse area to satisfy the adiabaticity and prevents in most cases
its use for pulses shorter than nanoseconds.

On the basis of the adiabatic theorem, it is commonly believed
that a large pulse area for adiabatic passage is inevitable and
that it is the price to pay to get the robustness. It has been
however found that, for a two-level system driven by an
appropriate shaped chirped pulse, the pulse area can be only
typically $1.5\pi$ \cite{Parallel}, which is not much larger than
the $\pi$ value (corresponding to the smallest value leading to a
complete population transfer for a pulse of a laser perfectly
tuned to the resonance, the $\pi$-pulse \cite{OptControl}). This
finding has been made from a geometric picture showing the
surfaces of eigenenergies as functions of the field parameters
\cite{Topology}: the paths that optimize the non-adiabatic losses
are level-lines in the diagram of the distance between the two
surfaces and correspond to instantaneous eigenvalues that are {\it
parallel} at all times.

In a $\Lambda$-system (where we assume that the two ground states
are not coupled, e.g. for symmetry reasons), the naive simplest
way to completely transfer the population between the two ground
states is two successive $\pi$ pulses, one for each transition,
leading to a total pulse area of $2\pi$. One can also consider two
exactly overlapping fields also giving a total pulse area of
$2\pi$ \cite{Shore}. It has been shown that the minimum area for
such a process is $\sqrt{3}\pi$, which corresponds to the
singular-Riemannian geodesic \cite{OptControl}.

We show in this paper that one can extend the optimization
procedure making the dynamics follow parallel eigenvalues at all
times to a $\Lambda$-system driven by two appropriately shaped
fields of chirped frequencies. One gets a complete population
transfer for a total pulse area as low as 3$\pi$ with the standard
robustness of adiabatic processes. This opens the possibility to
implement ultrafast Raman adiabatic passage technique, i.e. in the
pico-and subpicosecond regime, using the state-of-the-art
technology of shaped femtosecond laser pulses (see for instance
\cite{Weiner,Chatel} and its recent use and proposals for
strategies of control
\cite{Gerber,Silberberg,Brixner,Baumert,Ultimate}).

The paper is organized as follows. We first describe the technique
of parallel adiabatic passage in $\Lambda$-systems. We next
compare it with respect to the conventional STIRAP. Robustness
with respect to fluctuations is analyzed. Before concluding, we
propose an implementation of the parallel STIRAP using pulse
shaping techniques.

\section{Parallel adiabatic passage in $\Lambda$-systems}
\subsection{The system}
The $\Lambda$-system, with the two ground states denoted
respectively $|1\rangle$ and $|3\rangle$, driven by two fields,
respectively the pump and the Stokes fields, is described in the
resonant approximation \cite{Shore} in the dressed-state basis
%$\{|1\rangle,|2\rangle,|3\rangle\}$
by the Hamiltonian
\begin{equation}
{\sf H}^{[\Omega_P,\Omega_S,\Delta,\delta]}=\frac{1}{2}\left[
\begin{array}{ccc}
0 & \Omega_P & 0\\
\Omega_P &2\Delta & \Omega_S \\
0 & \Omega_S & 2\delta
\end{array}
\right]. \label{Model}
\end{equation}
The available parameters that can a priori vary during time are
$\Omega_P$ and $\Omega_S$ (chosen real for simplicity) the pump
and Stokes Rabi frequencies coupling the transitions
$|1\rangle-|2\rangle$ and $|3\rangle-|2\rangle$ respectively,
$\Delta$ and $\delta$ respectively the one-photon detuning (with
respect to the pump) and the two-photon detuning:
$\Delta=\omega_2-\omega_1-\omega_P$,
$\delta=\omega_3-\omega_1-\omega_P+\omega_S$ with $\hbar\omega_j$,
$j=1,2,3$, the energies of the corresponding state $|j\rangle$ and
$\omega_P$, $\omega_S$ the frequencies of the pump and Stokes
fields respectively.

\subsection{Construction of parallel eigenvalues}
We apply the analysis that was made in Ref \cite{Parallel} for a
two-state system to the three-state model: We require an adiabatic
passage process such that \emph{the eigenvalues stay parallel at
each time} in order to optimize the adiabatic passage. We first
remark that this imposes fields of chirped frequency. The use of
(non-optimized) linearly chirped fields has been proposed in
\cite{Band} and studied in \cite{Sola} with the aim of minimizing
the intermediate-level population. We first determine sufficient
conditions to achieve the parallelism of the eigenvalues and next
give a time-dependent realization that can be implemented using
modern pulse shapers.

Adiabatic passage processes require the knowledge of the
eigenelements. For the case of a two-photon detuning, they have
been determined and studied for instance in \cite{Fewell}. Here we
require the parallelism of the eigenvalues. More precisely,
denoting $\omega_{-,0,+}$ the three eigenvalues satisfying at each
time $\omega_-<\omega_0<\omega_+$ and $|\psi_{-,0,+}\rangle$ the
corresponding eigenstates, we impose for reasons of symmetry
$\omega_+-\omega_0=\omega_0-\omega_-$ and that $|\psi_{0}\rangle$
transports adiabatically the population, thus connecting at early
times the initially populated state $|1\rangle$ and at late times
the target state $|3\rangle$:
$|1\rangle\buildrel{t\to-\infty}\over\longleftarrow|\psi_0\rangle
\buildrel{t\to+\infty}\over\longrightarrow|3\rangle$. The
eigenvalues are thus of the form
\begin{equation}
\omega_0=\frac{1}{3}(\Delta+\delta),\quad
\omega_{\pm}=\frac{1}{3}(\Delta+\delta)\pm\frac{1}{2}\Omega_0,
\end{equation}
with
\begin{equation}
\label{Omega0}
\Omega_0=\sqrt{\Omega_P^2+\Omega_S^2+\frac{4}{3}(\Delta^2-\Delta\delta+\delta^2)}
\end{equation}
and the condition
\begin{equation}
\label{Cond}
0=[9(\Omega_P^2+\Omega_S^2)-4(2\Delta-\delta)(2\delta-\Delta)]\frac{\Delta+\delta}{27}
-\Omega_P^2\delta.
\end{equation}
The initial connection, when $\Omega_P=\Omega_S=0$, imposes for
the time-dependent detuning
$\delta(-\infty)=-\Delta(-\infty)=\Omega_0/2$ corresponding to
$\omega_0(-\infty)=0$, and $\omega_{\pm}(-\infty)=\pm\Omega_0/2$.
The final connection can be chosen in two ways: Either
$\omega_+(+\infty)=0$, $\omega_0(+\infty)=-\Omega_0/2$,
$\omega_-(+\infty)=-\Omega_0$, or $\omega_+(+\infty)=\Omega_0$,
$\omega_0(+\infty)=\Omega_0/2$, $\omega_-(+\infty)=0$. We have
chosen below the latter situation corresponding to
$\delta(+\infty)=\Omega_0/2$ and $\Delta(+\infty)=\Omega_0$. At
this stage, we expect many solutions for the four available
parameters from Eqs. \eqref{Omega0} and \eqref{Cond} with boundary
conditions:
$0\buildrel{t\to-\infty}\over\longleftarrow\Omega_{P,S}(t)
\buildrel{t\to+\infty}\over\longrightarrow0$,
$-\Omega_0/2\buildrel{t\to-\infty}\over\longleftarrow\Delta(t)
\buildrel{t\to+\infty}\over\longrightarrow\Omega_0$,
$\Omega_0/2\buildrel{t\to-\infty}\over\longleftarrow\delta(t)
\buildrel{t\to+\infty}\over\longrightarrow\Omega_0/2$.

\subsection{Numerical implementation}

\begin{center}
\begin{figure}[h]
\includegraphics[scale=0.6]{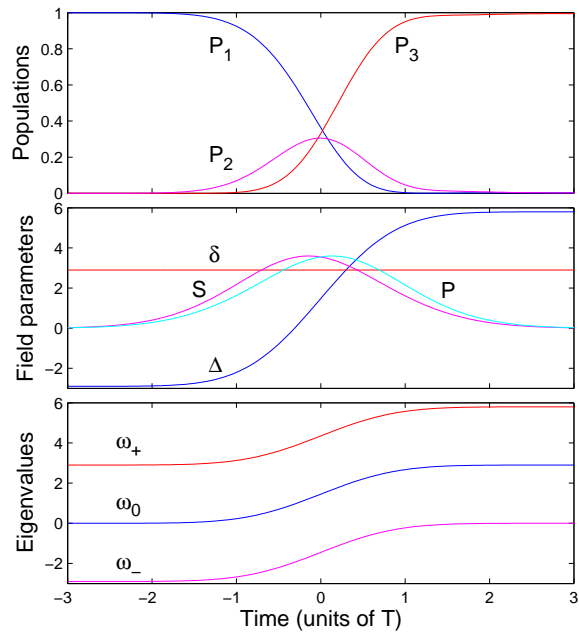}
\caption{Transition Probability for a level line dynamics as
function of time for $\Omega_0=5.8/T$ (corresponding to the total
pulse area ${\cal A}\approx3.6\pi$). Upper frame: Population
$P_j=|\langle j|\phi(t)\rangle|^2$ with $\phi(t)$ the state
solution. Middle frame: The field parameters (in units of $1/T$)
$\Delta(t)$, $\delta(t)$ given by Eqs. \eqref{Choice} (for
$\alpha=0$), and $\Omega_P(t)$ and $\Omega_S(t)$ determined from
Eqs. \eqref{Omega0} and \eqref{Cond}. Lower frame: The eigenvalues
(in units of $1/T$).} \label{Dyn_ll}
\end{figure}
\end{center}

To find a simple implementable solution, we additionally impose
the time parametrization as a monotonic smooth function for
$\Delta(t)$, symmetric around $t=0$, and a symmetric Gaussian
shape for $\delta(t)$:
\begin{equation}
\label{Choice} \Delta(t)=\frac{3\Omega_0}{4}
\text{erf}\bigl(\frac{t}{T}\bigr)+\frac{\Omega_0}{4},\
\delta(t)=\frac{\Omega_0}{2}[1+\alpha e^{-(\beta t/T)^2}],
\end{equation}
with $\alpha$ and $\beta$ to be chosen. The quantity $T$
corresponds to a characteristic time of the duration of the
process. The choice $\alpha=0$ is simple since it leads to a
constant two-photon detuning; it will be shown to give already
very good results. For given $\alpha$ and $\beta$, the choice
$\Omega_0$ comes down to choose a unique level line for the
problem and set the values of the Rabi frequencies $\Omega_P(t)$
and $\Omega_S(t)$ from Eqs. (3) and (4). Note that the
parametrization \eqref{Choice} is somewhat arbitrary. Other smooth
shapes for the detunings lead to very similar (but not strictly
identical) results.

We define the total pulse area for the process as
\begin{equation}
{\cal A
}=\int_{-\infty}^{+\infty}dt\sqrt{\Omega_P^2(t)+\Omega_S^2(t)}.
\end{equation}
This characteristic quantity is known to be at least $\pi$ when
one considers a two-state transition (for instance the 1-2
transition by the pump field) \cite{OptControl}. The pulse area
has a clear-cut dynamical meaning only for resonant pulses in
two-level systems. In other (nonresonant, multilevel) situations,
different pulses with the same individual area can produce
different results even in near-adiabatic cases. This choice of the
total pulse area takes also into account the delay between the
pulses. Considering for simplicity two identical delayed pulses of
individual area ${\cal A }_P$, the area ${\cal A}$ is bounded
between $\sqrt{2}{\cal A }_P$ (no delay) and $2{\cal A }_P$
(consecutive pulses without overlapping). This criterion favors
thus overlapping pulses since they lead to faster processes.
 In the $\Lambda$
system, the use of two successive $\pi$ pulses, respectively for
the pump and Stokes fields, or of two overlapping fields
corresponds both to ${\cal A}=2\pi$ \cite{Shore}. However it is
known that the minimum area for such a process is $\sqrt{3}\pi$
\cite{OptControl}. An alternative quantity of interest is the
total fluence corresponding to an integrated intensity:
\begin{equation}
{\cal F }=\int_{-\infty}^{+\infty}dt(\Omega_P^2(t)+\Omega_S^2(t)).
\end{equation}

Figure \ref{Dyn_ll} shows the resulting dynamics corresponding to
parallel eigenvalues with $\Omega_0=5.8/T$ and $\alpha=0$. We
obtain ${\cal A}\approx3.6\pi$ and ${\cal F}\approx40/T$ in these
conditions. The obtained final population transfer
$P_3(+\infty)\approx 0.995$ shows a very efficient population
transfer by adiabatic passage for a quite modest pulse area, only
approximately twice the minimum required area $\sqrt{3}\pi$. One
can notice the counterintuitive sequence of the pulses (i.e. the
Stokes field coming first before the pump field), which appears
here as a consequence of the level line procedure for the choice
\eqref{Choice}. We notice a non-negligible transient population in
the upper state 2 since here the dynamics is not governed by a
dark state. We however anticipate that a fast enough process,
typically subpicosecond, will allow one to neglect the resulting
loss from the upper state.

\section{Comparison with the conventional STIRAP}

\begin{center}
\begin{figure}[h]
\includegraphics[scale=0.58]{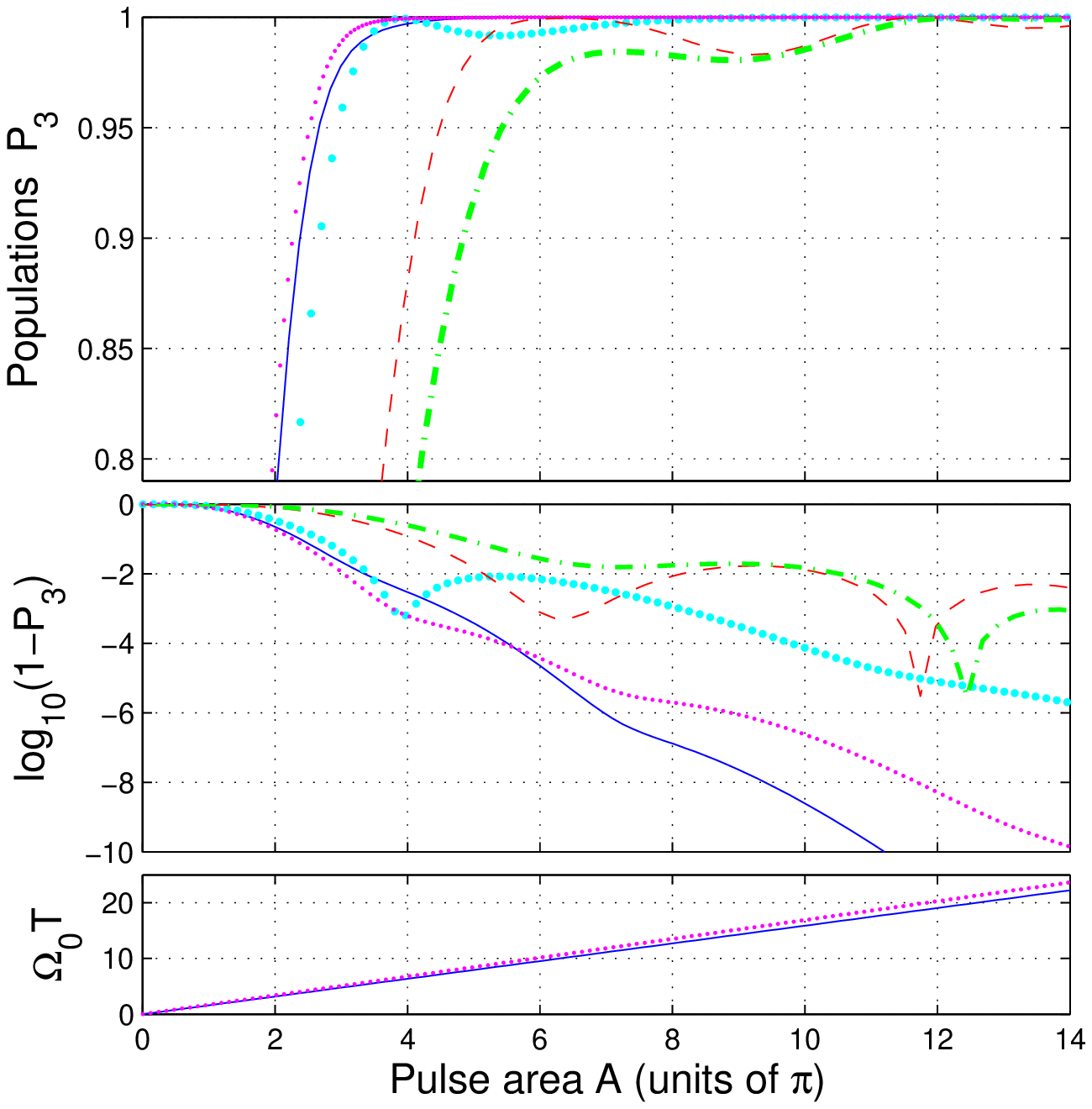}
\caption{Upper frame: Final population transfer for the parallel
STIRAP with $\alpha=0$ (full line), linearization of $\Delta(t)$
[see Eq. \eqref{Lineariz}] (large-dotted line),
%and with fitting pulse shapes
%\eqref{Fit} and linearization (at $t=0$) of $\Delta(t)$
%(large-dotted line)
the conventional STIRAP with the delay $\tau=1.1T$ (dashed line)
and with the delay $\tau=T$ (dot-dashed line), and parallel STIRAP
with $\alpha=0.1/T$ and $\beta=1.25$ (thin-dotted line) as a
function of the total pulse area ${\cal A}$. Middle frame:
logarithmic scale of the deviation from the complete population
transfer. Lower frame: The corresponding $\Omega_0T$ for the
parallel STIRAP's.} \label{S_S_ll_A}
%\end{figure}
%\end{center}
%
%\begin{center}
%\begin{figure}[h]
\includegraphics[scale=0.58]{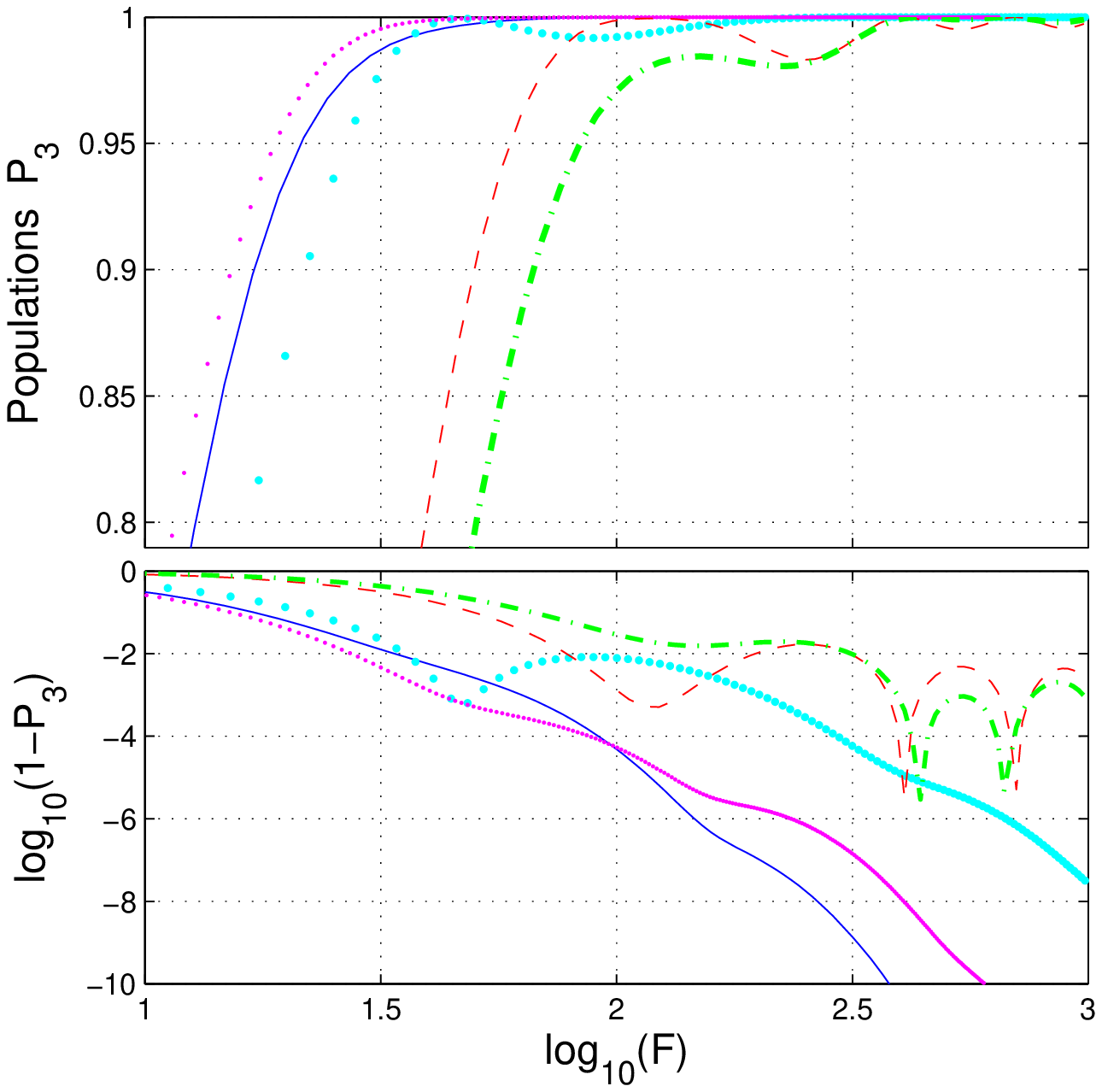}
\caption{Same as Fig. \ref{S_S_ll_A} but as a function of the
fluence ${\cal F}$ (in units of $1/T$) in logarithmic scale.}
\label{S_S_ll_F}
\end{figure}
\end{center}

Figures \ref{S_S_ll_A} and \ref{S_S_ll_F} show a comparison of the
Stimulated Raman parallel adiabatic passage using \eqref{Choice},
for a constant ($\alpha=0$) and a non-constant ($\alpha\ne0$)
two-photon detuning, with the conventional STIRAP in terms of
population transfer efficiency with respect to, respectively, the
pulse area ${\cal A}$ and the fluence ${\cal F}$. The conventional
STIRAP process is considered with a counterintuitive sequence of
Gaussian pulses of same peak values that are delayed of $\tau$:
$\Omega_P=\Omega_{\max}e^{-[(t-\tau/2)/T ]^2}$,
$\Omega_S=\Omega_{\max}e^{-[(t+\tau/2)/T]^2}$. Two values for
$\tau$ are considered (with a difference of 10\%) in Figs.
\ref{S_S_ll_A} and \ref{S_S_ll_F}: We can first notice a
significant dependence on this delay for the relatively small
areas that are here considered and thus a relative lack of
robustness. Note that we have also studied the influence of the
delay between the two pulses for the parallel STIRAP: We have
obtained a much less significant dependence.

For the constant two-photon detuning case ($\alpha=0$), the
population transfer to state 3 by the parallel strategy is already
very efficient for ${\cal A}\approx3.6\pi$ (corresponding to the
dynamics considered in Fig. \ref{Dyn_ll}) and gets monotonically
better for higher values of ${\cal A}$. This value ${\cal
A}\approx3.6\pi$ corresponds to a weak efficiency for the
conventional STIRAP. We recover this efficiency for the
conventional STIRAP for a pulse area approximately three times
larger than for the parallel strategy, corresponding to a fluence
approximately one order of magnitude larger (see Fig.
\ref{S_S_ll_F}). This shows the significant superiority of the
parallel strategy with respect to the conventional STIRAP. The
deviation from the complete population transfer displayed in Figs.
\ref{S_S_ll_A} and \ref{S_S_ll_F} with a logarithmic scale shows
the high efficiency of the parallel strategy. We obtain a transfer
with an accuracy to more than 4 digits for an area larger or equal
to only 5.5$\pi$. We remark that such accuracy for the population
transfer is not possible for the conventional STIRAP (see also
\cite{Vasilev}).

Figures \ref{S_S_ll_A} and \ref{S_S_ll_F} shows that the
efficiency of the parallel strategy can be improved using a
non-constant two-photon detuning ($\alpha\ne0$): We obtain the
same transfer (considering a deviation from the complete
population transfer not smaller than $10^{-5}$) with a significant
smaller area for the non-constant two-photon detuning case with
respect to the constant case. Numerics shows that the transfer is
generally better when $\alpha$ and $\beta$ are such that the two
pulses approximately overlap (requiring $\alpha\ne0$). The
transfer efficiency is shown in Fig. \ref{S_S_ll_A} in that case
for $\alpha=0.1\Omega_0$ and $\beta=1.25$. We obtain a very good
transfer efficiency already from ${\cal A}\approx 3\pi$. The gain
for the fluence with respect to the conventional STIRAP is more
than one order of magnitude as shown in Fig. \ref{S_S_ll_F}.

\section{Robustness of parallel STIRAP}

The robustness of the process with respect to an imperfect
knowledge of the pulse areas is shown in Fig. \ref{S_S_ll_A}.
Sources of other fluctuations can be various and the robustness of
the parallel strategy is questionable since it is mainly based on
the use of the Davis-Dykhne-Pechukas formula in the time complex
plane with analytic functions as the pulse parameters
\cite{Parallel}.

\begin{center}
\begin{figure}[h]
\includegraphics[scale=0.6]{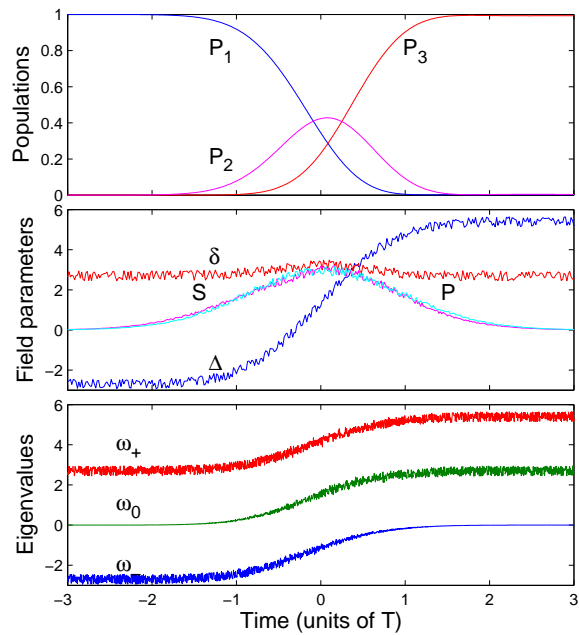}
\caption{Same as Fig. \ref{Dyn_ll} but for $\Omega_0=5.4/T$,
$\alpha=0.1$, and $\beta=1.25$, and with fluctuations for the
amplitudes and instantaneous frequencies. The field parameters and
eigenvalues are shown for a single realization.}
\label{Dyn_ll_Fluc}
\end{figure}
\end{center}

We consider a white noise of width $\Gamma=0.5T$ on the detunings
$\Delta_F(t)$ and $\delta_F(t)$, on the Rabi frequencies
$\Omega_{F,j}(t)$, $j=P,S$, and on their areas, defining
\begin{subequations}
\begin{eqnarray}
\label{NoiseAP} \Omega_{F,j}(t)&=&\Omega_j(t)+\Gamma \Lambda_j(t)(
r_1+r_2)\\
%\end{equation}
%\begin{equation}
%\label{NoiseP}
\Delta_F(t)&=&\Delta(t)+\Gamma r_3, \quad
\delta_F(t)=\delta(t)+\Gamma r_4
\end{eqnarray}
\end{subequations}
with $\Omega_j(t)$ the Rabi frequency of the field $j$ without
fluctuation, $0\le\Lambda_j(t)\le1$ the shape of the field $j$
without fluctuation, $\Delta(t)$ and $\delta(t)$ the instantaneous
detunings without fluctuation, and $-0.5\le r_k< 0.5$,
$k=1,\cdots,4$, four (independent) random numbers. The random
number $r_1$ is drawn once for each realization of a complete
dynamics. It allows one to vary randomly the pulse areas. The
random numbers $r_k$, $k\ge2$ have been drawn at each time step of
calculation (here taken as $T/300$) for all realizations. This
allows one to model instantaneous fluctuations of the detunings
and of the pulse shapes. We have determined the instantaneous
populations shown in Fig. \ref{Dyn_ll_Fluc} by averaging over many
realizations of time histories (Monte Carlo simulation). The
averaged final population transfer is here $P_3(+\infty)\approx
0.993$ (to be compared to $P_3(+\infty)\approx 0.996$ without
fluctuations).

% Figure \ref{Robust_Fig} investigates
%the robustness of the parallel strategy with respect to noise, on
%field amplitude and detunings, for various noise width
%$\Gamma$: We obtain a very good robustness for quite large widths.
%Figure \ref{Robust_Fig} also compares the robustness of the
%parallel strategy with the standard STIRAP such that the transfer
%efficiency is the same without fluctuation (corresponding to a
%total field area of approximately three times larger for the
%standard STIRAP.) It shows that the robustness of the two
%processes are comparable.
%%To recover the same efficiency, one needs
%%to increase the pulse area to ${\cal A}\approx12$.
%
%\begin{center}
%\begin{figure}[h]
%%\includegraphics[scale=0.6]{Fluct.eps}
%\caption{Final population transfer for the parallel STIRAP (full
%line) with $\alpha=0$ and $\Omega_0T=5.8$ (leading to ${\cal
%A}\approx3.6$) and for standard STIRAP (dashed line) with
%$\Omega_{\max}T=12$ (leading to ${\cal A}\approx11.2$) as a
%function of the noise width $\Gamma$ applied solely on the
%amplitude (upper full and dashed lines), and applied on both the
%detunings and field amplitudes (lower full and dashed lines).}
%\label{Robust_Fig}
%\end{figure}
%\end{center}

\section{Implementation of parallel STIRAP by pulse shaping techniques}

\begin{center}
\begin{figure}[h]
\includegraphics[scale=0.7]{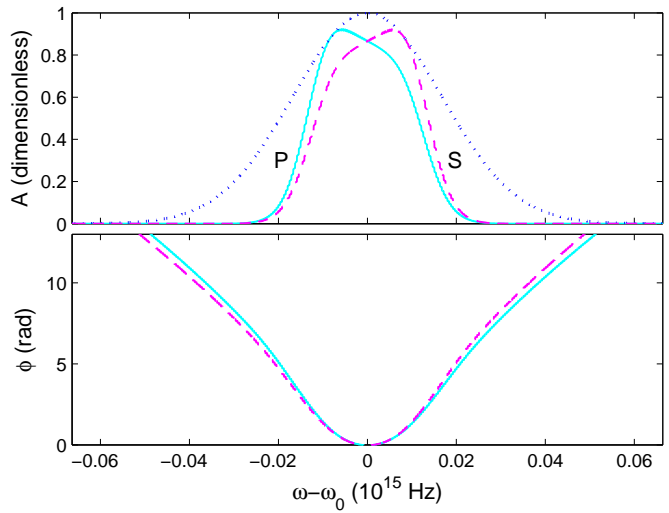}
\caption{Tailored spectral normalized amplitudes $A$ (upper frame)
and phases (lower frame) for the pump (full lines) and Stokes
(dashed lines) fields from 100 fs Fourier transform limited pulses
(upper frame, dotted line) of mean frequency $\omega_0$ to produce
the fields of Fig. \ref{Dyn_ll}.} \label{Shaping_Fig}
\end{figure}
\end{center}

The implementation of the parallel STIRAP requires the design of
the pulses and frequencies using the state-of-the-art technology
of pulse shaping, such as for instance liquid cristal spatial
light modulators. Two such modulators allow the independent
shaping of the spectral phase and amplitude of an input
femtosecond laser pulse \cite{Weiner,Chatel}. We consider here for
simplicity a constant two-photon detuning ($\alpha=0$). In that
case, we have to chirp both field frequencies exactly in the same
way. The difference between the two fields is their mean frequency
and their delayed envelope. We can determine approximate curve
fitting of the envelopes for $\alpha=0$:
\begin{equation} \label{Fit}
\Omega_{P,S}(t)\approx\frac{\Omega_0}{2}\beta
e^{-[(t\pm0.14T)/(\beta T)]^2},\quad \beta=\sqrt\frac{\pi}{2}
\end{equation}
with $+$ ($-$) holding respectively for the Stokes (pump) field.
%These fitting lead to the same efficiency of the one shown in Fig.
%\ref{S_S_ll} (upper frame, full line).
Figure \ref{Shaping_Fig} shows, for these fields of phase
corresponding to \eqref{Choice} with $\alpha=0$ (cf. conditions of
Fig. 1), their corresponding spectral amplitudes and phases,
respectively denoted $A(\omega)$ and $\phi(\omega)$, and defined
through the complex field as ${\cal
E}\Lambda(t)e^{i\varphi(t)}=\frac{1}{2\pi}\int d\omega
A(\omega)e^{i[\omega t+\phi(\omega)]}$ (for each field). Here
$\varphi(t)$ is the instantaneous phase, leading the instantaneous
frequency $d\varphi/dt$, $0\le\Lambda(t)\le1$ the shape of the
field amplitude, and ${\cal E}$ the peak amplitude of the
considered field. The instantaneous frequencies of the pump field
$\omega_P\equiv d\varphi_P/dt$ and of the Stokes field
$\omega_S\equiv d\varphi_S/dt$ are connected to the one- and
two-photon detunings as follows:
$\omega_P(t)=\omega_2-\omega_1-\Delta(t)$,
$\omega_S(t)=\omega_2-\omega_3+\delta-\Delta(t)$. We have chosen
for convenience the mean frequency of the pump as
$\omega_{0,P}=\omega_2-\omega_1-\Omega_0/4\equiv \omega_P(0)$ and
of the Stokes as
$\omega_{0,S}=\omega_2-\omega_3+\delta-\Omega_0/4\equiv
\omega_S(0)$. We have considered fields of full width at half
maximum $T_{\text{FWHM}}=500$ fs corresponding to $\Omega_0=17.1$
THz for the considered case $\Omega_0=5.8/T$. For typical
transition strengths of 1 Debye, this leads to fields of peak
intensities 15 GW/cm$^2$. For each field, the spectral parameters
of Fig. \ref{Shaping_Fig} have been generated from a Gaussian
Fourier transform limited pulse centered at the mean frequency
$\omega_0$ with a full width at half maximum of 100 fs, shaped in
phase and amplitude using two spatial light modulators. Note that
we have considered the realistic effect of pixelization for the
modulators (using 320 pixels, almost not noticeable at the scale
of Fig. \ref{Shaping_Fig}) which does not show any noticeable
difference for the output pulses and the corresponding dynamics of
Fig. \ref{Dyn_ll}.

For practical implementation, it could be of interest to
alternatively use, with the fitting Gaussian pulses \eqref{Fit},
linearly chirped frequencies constructed from the {\it
linearization} of \eqref{Choice} around $t=0$:
\begin{equation}
\label{Lineariz} \Delta(t)= \frac{3\Omega_0}{2T\sqrt{\pi}}t,
\end{equation}
 where
$\sqrt{\Omega_P^2+\Omega_S^2}$ is maximum. This would indeed allow
one to use for instance gratings to produce the chirp instead of
modulators (see for instance \cite{Noordam}). In this case, the
duration of the pulse is connected to the slope of the chirp and
the pulses have to be delayed additionally. Its efficiency is
shown in Figs. \ref{S_S_ll_A} and \ref{S_S_ll_F}: It is not as
efficient as the level line strategy but is still well superior to
the standard STIRAP. For instance, one can see that the deviation
from the complete transfer is less than $10^{-2}$ with a pulse
area larger than 3.5$\pi$ for this linearization, corresponding to
a fluence larger than $10^{1.6}/T$. One recovers this accuracy for
STIRAP for a fluence larger than 10.5$\pi$, corresponding to a
fluence larger than $10^{2.5}/T$ (almost one order of magnitude
larger). This linearization also allows an error smaller than
10$^{-4}$ for an area larger than 10$\pi$ (it should be larger
than only 5.5$\pi$ for parallel STIRAP). Such accuracy is not
reachable using the conventional STIRAP.

\section{Conclusion and discussions}

%Sources of fluctuations in the process are of very different
%nature, when comparing the standard STIRAP process usually
%implemented with nanosecond lasers and the parallel STIRAP
%proposed here by shaped femtosecond lasers. For the nanosecond
%pulses, the main source of noise comes from the phase fluctuations
%(see for instance \cite{yatsenko}). For the shaped femtosecond
%pulses, that are mode-locked, the source of fluctuation
%essentially comes from the shape of the input pulse and the
%pixelization of the modulators \cite{Pixel}.

%We have analyzed
%the resulting efficiency when considering the pixellisation of the
%modulators (see also Fig. \ref{Shaping} to see an example of such
%pixellisation).

In conclusion, we have developed a strategy of adiabatic passage
in $\Lambda$-systems with field designed such that the eigenvalues
stay parallel at each time. This strategy allows one to combine
the energetically efficiency of ``$\pi$-pulse'' and related
strategies with the robustness of adiabatic passage. It has been
shown in particular to be superior to the conventional STIRAP. It
is expected to be easily implementable using modern tools of pulse
shaping.

To implement the proposed strategy in real systems, one has to
consider several limitations due to the presence of other states
and to lossy processes such as ionization. In order to avoid the
latter, the requirements are to consider a system which does not
show a one-photon resonance into the continuum from the excited
state and to use moderate field intensities, typically not more
than 100 GW/cm$^2$ for atoms. Additional non-resonant states lead
to Stark shifts which can be considered in general negligible with
respect to the one-photon resonances considered here. However for
the considered field intensities, they can play a role in
deforming the dressed energies. In that case one has to
recalculate the field parameters such that the eigenenergies stay
parallel. The conditions that permit these additional states to
stay non resonant can be estimated as $\Delta^{(j)}_k T\gg 1$ and
$\Omega_k^{(j)}\lesssim \Delta^{(j)}_k$ with $\Delta^{(j)}_k$ the
one-photon detuning for the state $j$ and the field $k$ and
$\Omega_k^{(j)}$ the corresponding peak Rabi frequency.

This parallel adiabatic passage strategy should be adapted to
produce in an optimal way superpositions of states such as in the
case of fractional STIRAP \cite{fstirap}. This should find
applications in quantum information processing, for instance to
implement fast quantum gates or quantum algorithm (such as the
implementation of the parallel adiabatic passage for the quantum
search \cite{QS}). The high efficiency of the parallel strategy
shown in Fig. 2 is compatible with such applications imposing
severe limitations on the admissible error, typically on a level
below $10^{-4}$ (see, e.g., \cite{Vasilev} and references
therein).

%\textit{Note.} While completing this work, we have been pointed
%out the very recent Reference \cite{Vasilev} that also deals with
%optimization of the conventional STIRAP. This reference limits its
%study to the shaping of the amplitude of the pulses.

%\vfill

\begin{acknowledgments}
We acknowledge the support from the French Agence Nationale de la
Recherche (projet CoMoC), the European Commission project
FASTQUAST, and from the Conseil R\'egional de Bourgogne. S.G.
thanks E. Hertz for fruitful discussions.
\end{acknowledgments}

\end{document}